# Experimental feature of bicycle flow and its modeling

Rui Jiang[1,2,3], Mao-Bin Hu[2], Qing-Song Wu[2], Wei-Guo Song[3]

[1] MOE Key Laboratory for Urban Transportation Complex Systems Theory and Technology, Beijing Jiaotong University, Beijing 100044, China

[2] School of Engineering Science, University of Science and Technology of China, Hefei 230026, China

[3] State Key Laboratory of Fire Science, University of Science and Technology of China, Hefei 230026, China

As a healthy and green transportation mode, cycling has been advocated by many governments. However, compared with vast amounts of works on vehicle flow and pedestrian flow, the bicycle flow study currently lags far behind. We have carried out experimental studies on bicycle flow on a 146-m long circular road. We present the fundamental diagram of bicycle flow and the trajectories of each bicycle. We have analyzed the spatiotemporal evolution of bicycle flow as well as its stability and phase transition behavior. The similarity between bicycle flow and vehicle flow has been discussed. We have proposed a cellular automaton model of bicycle flow, and the simulation results are in good agreement with the experiments.

Due to serious congestion and pollution, the traffic problem caused by motor vehicles has become an urban illness. As an ecological means of transportation, cycling has a number of advantages over other modes, both for the individuals and the society. For individuals, cycling is healthy and cheap, and sometimes can be faster than other transport modes. For society, cycling is environmental sustainable, requires cheap infrastructure and improves public health. Therefore, cycling has been advocated by many governments. On the one hand, extensive cycling rights of way have been complemented by many pro-bicycle policies and programmes such as ample bicycle parking, full integration with public transport. On the other hand, car parking has been made expensive and inconvenient in central cities, and strict land-use policies have generated shorter and thus more pro-cycling trips. As a consequence, the percent of bicycle trips has reached nontrivial level in some countries, such as 10% in Germany and Sweden, 11% in Finland, 18% in Denmark, and 27% in the Netherlands [1]. Therefore, bicycle flow study is practically important for design, operation and control of bicycle facilities.

On the other hand, as vehicle flow and pedestrian flow, bicycle flow is a self-driven many-particle system far from equilibrium [2,3]. Thus the study is expected to contribute to understand various fundamental aspects of nonequilibrium systems which are of current interest in statistical physics such as swarming and herding of birds and fishes [4], traveling waves in an emperor penguin huddle [5], ant trails [6], and movement of molecular motors along filaments [7].

However, unfortunately, compared with vast amounts of papers studying motor vehicle flow and pedestrian flow [2,3,8], the state of knowledge regarding bicycle flow currently lags far behind [9]. In the bicycle studies, the data were collected via on-site observations [10-15] and experiments [15]. The objective of these studies was to determine the optimal bike path or lane width, to assess safety, the impact of bicycles on vehicle traffic flow, the level of service, the fundamental diagram and the capacity of bicycle flow [9-20]. For example, the studies, so far, suggest that capacity varies in a wide range, such as 2,600 bicycles per hour per 3.3-ft (1 m) [11], 3,000 to 3,500 bicycles per hour per 2.6-ft (0.78 m) [12], 4,500 bicycles per hour per 8-ft (2.4 m) [13], 10, 000 bicycles per hour per 2.5-m [15].

Nevertheless, none of the previous studies investigates the phase transition feature of bicycle flow. This paper carries out an experimental study on the bicycle flow, see section Methods. We present not only the fundamental diagram, but also the trajectories of each bicycle so that we can analyze the spatiotemporal evolution of bicycle flow as well as its stability and phase transition behavior. We have also discussed the similarity between bicycle flow and vehicle flow, and presented a model to simulate the bicycle flow.

**Results**

Figure 1 shows the fundamental diagram of flow rate versus bicycle number, which exhibits the common feature as vehicular flow and pedestrian flow: the flow rate increases when bicycle density (bicycle number/course length) is small and decreases when bicycle density is large. The capacity is reached in the intermediate density.

For the first experiment, the capacity is estimated to be around 3000 bicycles per hour, complemented with the simulation (see section Discussion). For the second experiment, the capacity is a little smaller, which is around 2700 bicycles per hour. This may be due to the different gender constitution of the riders as pointed out in section Methods, and the maximum riding speed of male students is larger than that of female students. In the third experiment, flow rate is remarkably smaller due to the rain.

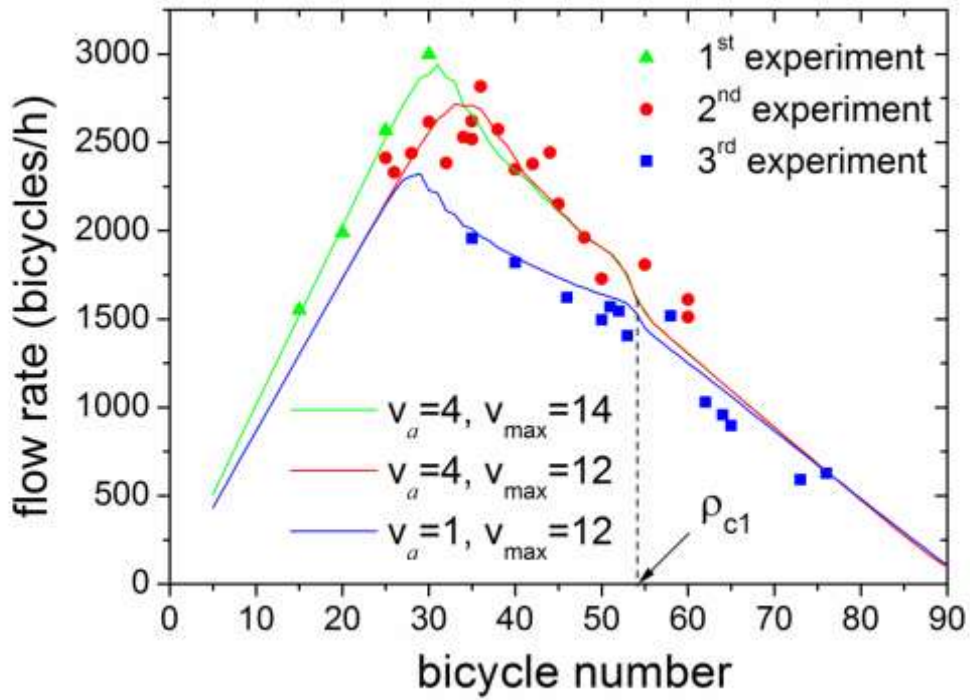

**Figure 1 The fundamental diagram obtained from the experiment (scattered data) and from the simulation (solid lines).**

Next we study the spatiotemporal evolution of the bicycle flow via investigating the trajectories of the bicycles. It has been found that there exists a critical density $\rho_{c1}$ around 37 bicycles per 100-m. Below the critical density, the bicycle flow is quite stable. Fig.2(a) and (b) show two such typical trajectory diagrams, see also the Supplementary Videos S1 and S2. One can see that even if small jams have occurred from time to time, they will soon disappear, as shown by the blue arrows. Moreover, as shown by the red arrows in Fig.2(a), some large gaps occur from time to time when the density is far below $\rho_{c1}$. On the one hand, this is because, when the preceding bicycle is slow, some riders would rather ride slowly instead of accelerating to decrease the gap and then decelerating. On the other hand, the behavior may be partially due to the rain. As a result, the flow rate is remarkably smaller. As the density approaches $\rho_{c1}$, the bicycle flow becomes more congested and the large gaps cannot occur.

However, when the density is above $\rho_{c1}$, the bicycle flow becomes unstable as shown in Fig.3(c). Traffic jams exist in the bicycle flow, see also Supplementary Video S3, from which one can see that some riders walked instead of riding even if outside of the jams because the riding speed approximately equals to walking speed. Even if the jam occasionally dissipates (in the vicinity of 700 s as shown by the blue arrow), it soon appears spontaneously (in the vicinity of 800 s as shown by the red arrow). When the density further increases, the jams cannot dissipate.

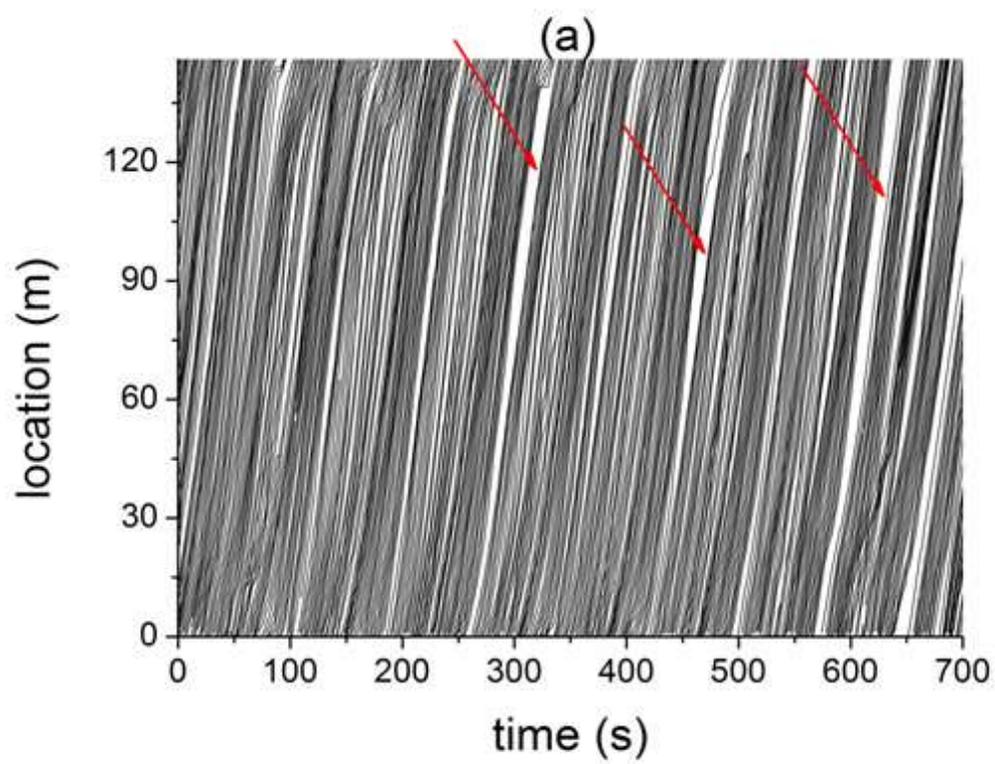

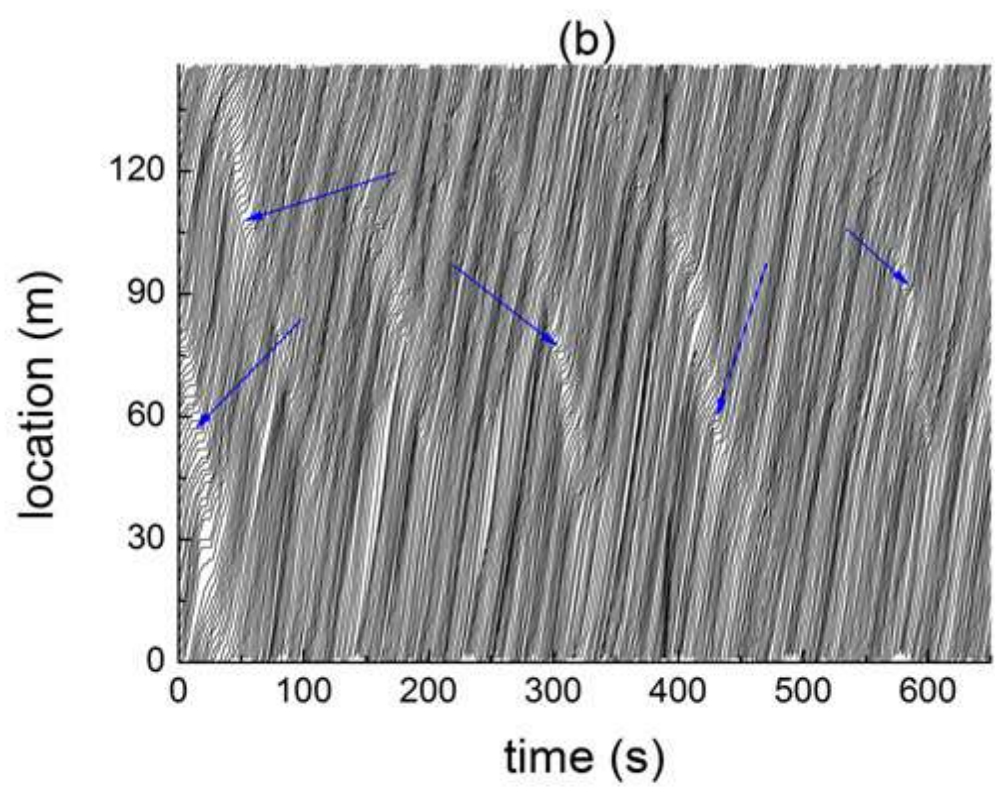

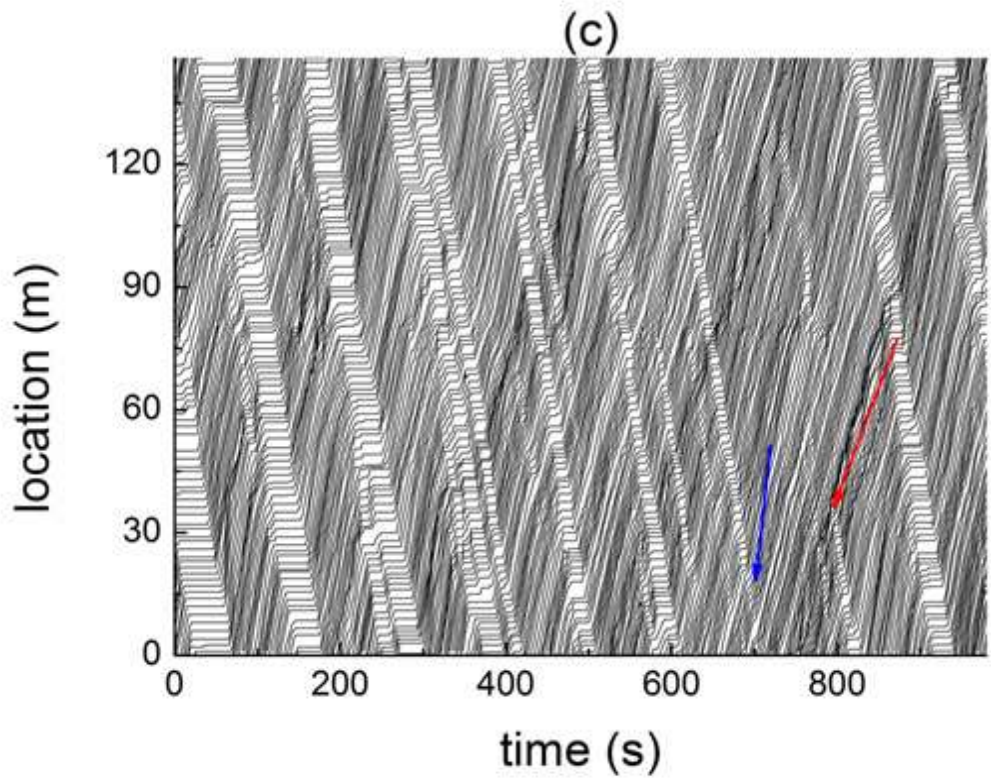

**Figure 2 Typical trajectory diagrams from experiment.** (a) from the third experiment and bicycle number is 39; (b) from the second experiment and bicycle number is 48; (c) from the third experiment and bicycle number is 63.

Recent empirical studies show that the congested vehicle flow can be classified into two phases: the synchronized flow and the jam [21-23]. When the density of synchronized flow is large (which is named heavy synchronized flow), the vehicle flow is not stable and a spontaneous phase transition from synchronized flow to jam will occur [21-26]. On the other hand, the light synchronized flow is quite stable and no spontaneous phase transition to jam occurs, and vehicle flow evolves into a widening synchronized flow pattern at the road bottlenecks [21-23]. One can see that the feature of bicycle flow is very similar to that of vehicle flow: the heavy (light) synchronized flow corresponds to bicycle flow with density above (below) $\rho_{c1}$.

In the future work, we need to carry out experimental study on wider courses allowing overtaking. Moreover, the effect of bottlenecks, which will cause phase transition from free bicycle flow to congested bicycle flow, also needs to be investigated. Based on the similarity between bicycle flow and vehicle flow, we expect that the experiment on bicycle flow will be helpful to disclose the mechanism of phase transition from free vehicle flow to congested vehicle flow.

**Discussion**

*Model*

Now we present a cellular automaton (CA) model of bicycle flow. The CA approach has been used successfully to simulate vehicular flow and pedestrian flow [27-30]. In the CA model, the road is classified into cells, and bicycles move with integer velocity 0, 1, ... , $v_{max}$. Here $v_{max}$ denotes the maximum velocity of bicycles. In each time step, the bicycles move as follows:

(1) Acceleration, $v_i$ → min($v_i$ +1, $v_{max}$)
(2) Deceleration,
    (2a) if $d_i < d_{od}$,   $v_i$ → min($v_i$, $d_i$, max($d_{i-1}$, $d_c$))
    (2b) else,         $v_i$ → min($v_i$, $d_i$)
(3) Determination of virtual velocity of a bicycle, $v_{vir,i}$ = max($v_i - 1, 0$)
(4) Recalculation of velocity, taking into account virtual velocity of preceding bicycle,
    $v_i$ → min($v_i$ + min($v_{vir,i-1}$, $v_a$), $v_{max}$)
(5) Randomization, $v_i$ → max($v_i - 1, 0$) with probability $p$
(6) Bicycle movement, $x_i$ → $x_i + v_i$

In the model, $v_i$ and $x_i$ denote velocity and position of bicycle $i$, respectively, $d_i = x_{i-1} - x_i - l$ is the gap between bicycle $i$ and its preceding one $i - 1$, $l$ is length of a bicycle.

The anticipation effect is considered in deceleration rule (2a), i.e., when a bicycle and its preceding one are within an operating distance $d_{od}$, the bicycle's velocity is decided not only by its spatial gap, but also by its preceding one's spatial gap. This reflects the fact that when the spatial gap of its preceding car is small, bicycle tends to move more slowly than that allowed by its spatial gap in order to avoid unrealistic oversized deceleration in the next time step. The parameter $d_c$ is a criterion parameter which indicates that the preceding bicycle's spatial gap has limited effect on the bicycle compared with its own spatial gap. For instance, even if the preceding bicycle's spatial gap $d_{i-1} = 0$, a bicycle still moves forward if its own gap $d_i > 0$. In rules (3) and (4), the virtual velocity of the preceding bicycle is introduced as in vehicular flow models [30], which takes into account the fact that the preceding bicycle is also moving. Here $v_a$ denotes the maximum limit of virtual velocity.

For the randomization probability $p$, the slow-to-start rule is taken into account by adopting a velocity dependent randomization probability as in vehicle flow model [28]:

$$p = \begin{cases} p_0 \\ p_n \end{cases} \quad \text{if} \quad \begin{matrix} v_i = 0 \\ v_i > 0 \end{matrix}$$

with $p_0 > p_n$.

*Simulation results*

In the simulations, the circular road is classified into 486 cells and each cell corresponds to 0.3 m. Thus, the road length is 145.8 m. The length of bicycles is set as 1.5 m, thus a bicycle occupies 5 cells. The parameters $p_n$ = 0.3, $p_0$ = 0.8, $d_c$ = 3 cells, and $d_{od}$ = 20 cells are used.

We compare the simulation results with the experimental ones. For the first set of simulation, the parameters are set as $v_{max}$ = 14, $v_a$ = 4, because the riders are all young male university student.

For the second set of simulation, the parameters are set as $v_{max}$ = 12, $v_a$ = 4 because there are both male and female riders. For the third set of simulation, the parameters are set as $v_{max}$ = 12, $v_a$ = 1 due to the rain. As can be seen from Fig.1, the simulation results are in good agreement with the experimental ones.

Figure 3 shows typical trajectory diagrams of the bicycle flow from the simulation. When the density $\rho < \rho_{c1}$, traffic jam disappears quickly and bicycle flow becomes homogeneous even if starting from a standing platoon, see Fig.3(a) and (b). On the other hand, when the density is larger than $\rho_{c1}$, the traffic jam cannot dissipate, see Fig.3(c). This is also generally in consistent with the experiment.

The deficiency of the model is that the bicycle flow is quite homogeneous. Therefore, it cannot reproduce the spontaneous formation and disappearance of small jams shown in Fig.2(b), and the disappearance and re-formation of jam shown in Fig.2(c). Thus, the model needs to be improved in the future work.

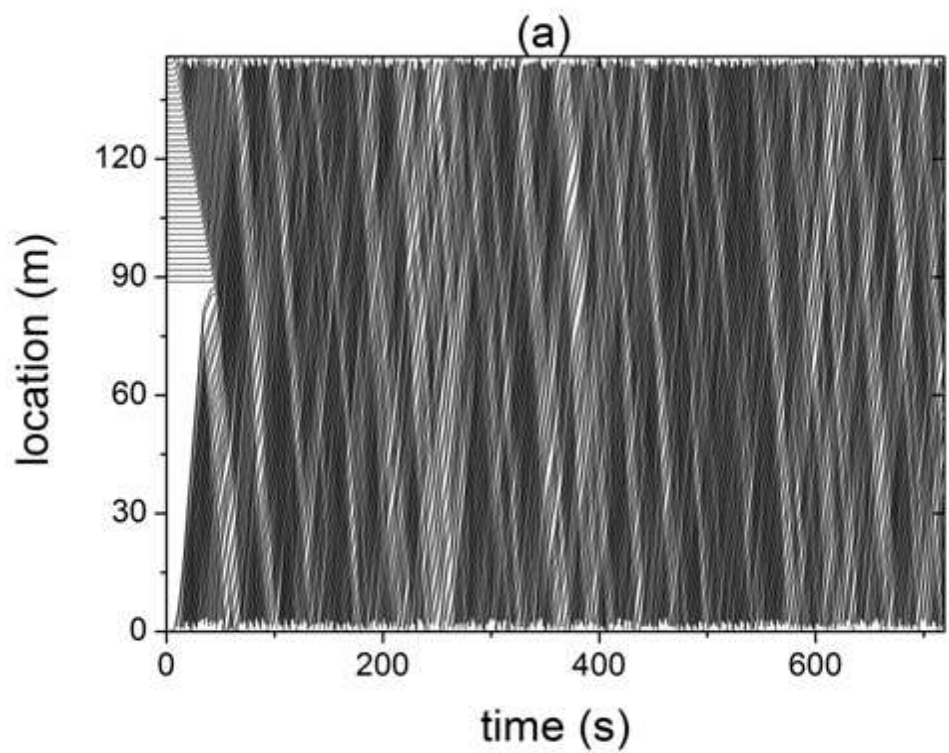

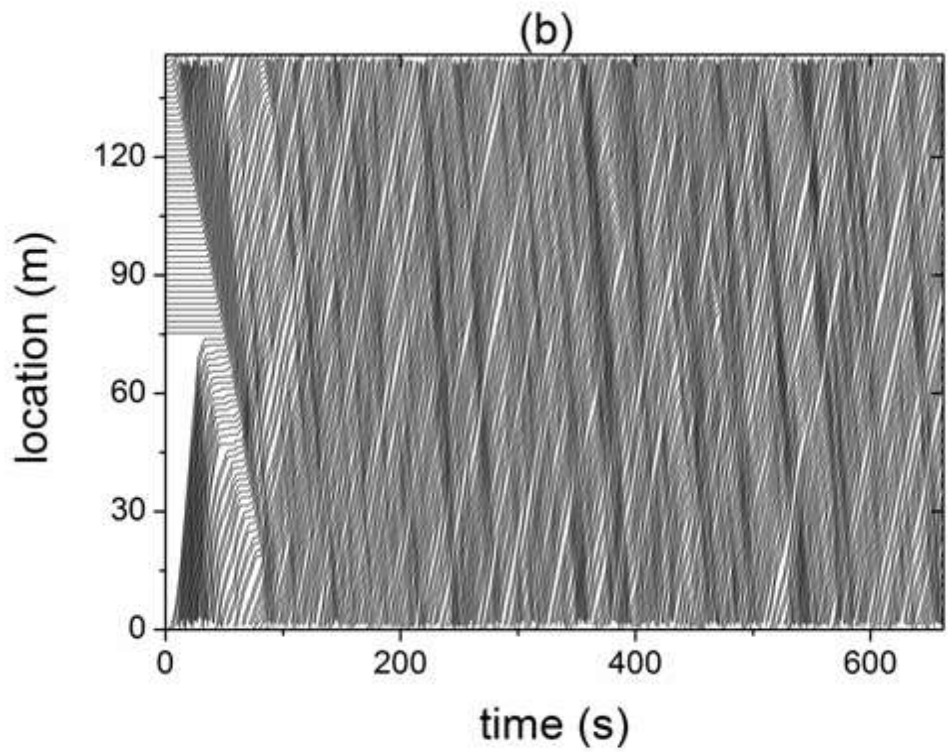

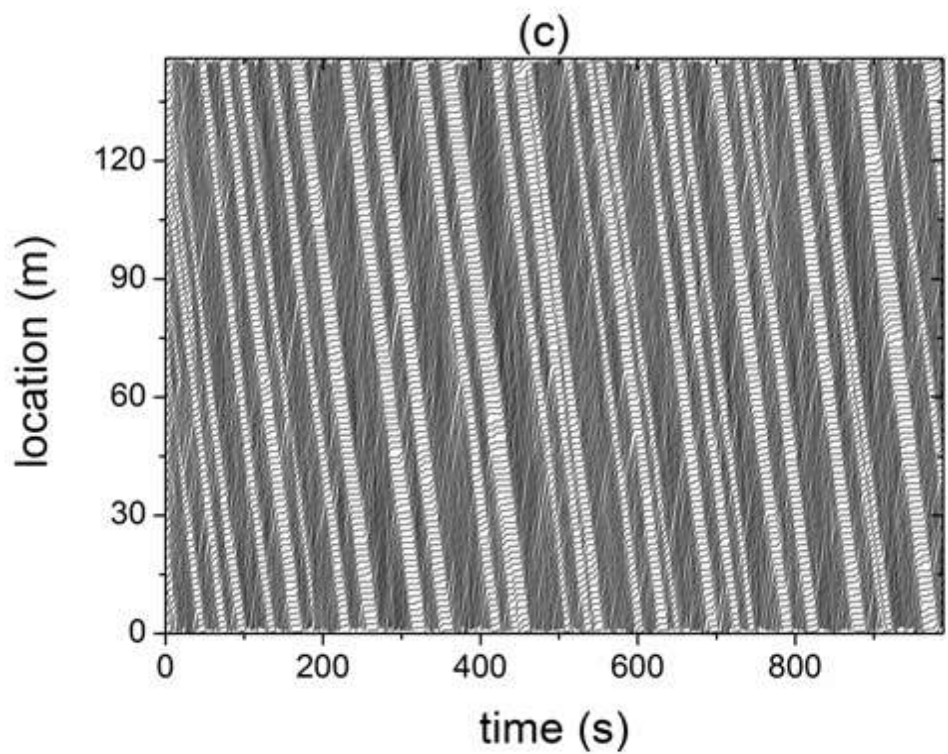

**Figure 3 Typical trajectory diagrams from simulation.** (a) the third set of parameters and bicycle number is 39; (b) the second set of parameters and bicycle number is 48; (c) the third set of parameters and bicycle number is 63.

**Methods**

In the experiment, we use an oval-like course on four neighboring basketball court with 29 m straight sections joined by 14 m circular curves, see Fig.4. The total length of the course is thus about 146 m. The width of the course is about 80 cm and bicycles are not allowed to overtake. We have changed the number of bicycles within the course and thus observed features of bicycle flow under different density. Note that this kind of experiment has already been conducted for vehicle flow and pedestrian flow [24-26, 31].

We have conducted the experiment three times, on Nov.3.2009, Nov.7.2009, Apr.19.2010, respectively. The weather is sunny for the first time, sunny/cloudy for the second time, and a little rainy for the third time. Thus some riders hold up an umbrella with one hand while riding in the third experiment. In the first experiment, the riders are all young male university student. In the second and third experiments, there are both male and female riders. A video camera has been used to film the experiment on a 18-stories building neighboring the experiment field.

We have labeled the course into meters, via which the location information of each bicycle has been obtained every second from the videos. Thus, the trajectories of each bicycle can be plotted. For the flow, we counted the number of bicycles passing a point in unit time.

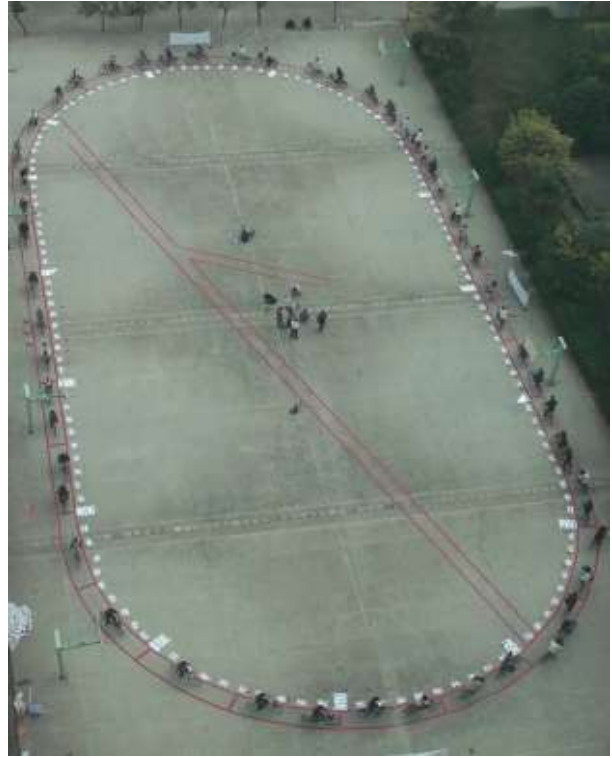

**Figure 4 A snapshot of the experiment**

**Supplementary information**

Video S1     Scenario of the experiment corresponding to Fig.2(a).
Video S2     Scenario of the experiment corresponding to Fig.2(b).
Video S3     Scenario of the experiment corresponding to Fig.2(c).

**Acknowledgments**

This work is funded by the National Basic Research Program of China (No.2012CB725404), the National Natural Science Foundation of China (Grant Nos. 71371175, 11422221 and 71171185).

**Author contributions**

R.J. conceived and designed the research, R.J., M.B.H., Q.S.W., W.G.S. performed the experiment, R.J. analyzed the data and wrote the paper, all authors discussed the results and commented on the manuscript.

**Additional information**

Competing financial interests: The authors declare no competing financial interests.